\documentclass[aps,twocolumn,showpacs]{revtex4}
\usepackage{graphicx}
\usepackage{amsmath} 
\usepackage{amssymb}

\newcommand{\rhocluster}{\rho_{\rm pc}}

\newcommand{\myext}{eps}

\newcommand{\ec}{\eta_c}
\newcommand{\ep}{\eta_p}

\newcommand{\rv}{{\bf r}}

\newcommand{\eal}{{\bf e}_{\alpha}}
\newcommand{\crit}{{\rm crit}}
\newcommand{\spin}{{\rm spin}}

\begin{document}

\date{\today}

\author{Jos\'e A.\ Cuesta}
\author{Luis Lafuente}
\affiliation{
 Grupo Interdisciplinar de Sistemas Complejos (GISC),
 Departamento de Matem\'aticas, Universidad Carlos III de Madrid, Avenida de la
 Universidad 30, E-28911 Legan\'es, Madrid, Spain
}

\author{Matthias Schmidt}
\affiliation{
  Institut f{\"u}r Theoretische Physik II,
  Heinrich-Heine-Universit{\"a}t D{\"u}sseldorf,
  Universit{\"a}tsstra{\ss}e 1, D-40225 D{\"u}sseldorf, Germany
}

\title{Lattice density functional for colloid-polymer mixtures:\\
  Multi-occupancy versus Highlander version}

\begin{abstract}
  We consider a binary mixture of colloid and polymer particles with
  positions on a simple cubic lattice. Colloids exclude both colloids
  and polymers from nearest neighbor sites. Polymers are treated as
  effective particles that are mutually non-interacting, but exclude
  colloids from neighboring sites; this is a discrete version of the
  (continuum) Asakura-Oosawa-Vrij model. Two alternative density
  functionals are proposed and compared in detail. The first is based
  on multi-occupancy in the zero-dimensional limit of the bare model,
  analogous to the corresponding continuum theory that reproduces the
  bulk fluid free energy of free volume theory. The second is based on
  mapping the polymers onto a multicomponent mixture of polymer
  clusters that are shown to behave as hard cores; the corresponding
  (Highlander) property of the extended model in strong confinement
  permits direct treatment with lattice fundamental measure
  theory. Both theories predict the same topology for the phase
  diagram with a continuous fluid-fcc freezing transition at low
  polymer fugacity and, upon crossing a tricritical point, a
  first-order freezing transition for high polymer fugacities with
  rapidly broadening density jump.
\end{abstract}



\maketitle

\section{Introduction}

Mixtures of colloidal particles and non-adsorbing polymers suspended
in a common solvent \cite{poon02,tuinier03review} have been
theoretically investigated on various levels of description, ranging
from simplistic to realistic effective interactions between the
constituent particles. The prototype of the former is the
Asakura-Oosawa-Vrij (AOV) model \cite{asakura54,vrij76} that describes
the colloids as hard spheres and the polymers as ideal (i.e.,
non-interacting) effective spheres that interact via hard core
repulsion with the colloids.  Despite its simplicity this model
reproduces the essential trends in bulk phase behavior of
colloid-polymer mixtures, involving colloidal gas, liquid, and
crystalline phases
\cite{lekkerkerker92,meijer94,dijkstra99,bolhuis02phasediag}, and
proved to be useful for studying a range of interfacial properties
\cite{brader03swetl}, like the structure of colloidal gas-liquid
interfaces and wetting of substrates, issues that are experimentally
relevant \cite{aarts03swet,aarts04codef,wijting03}.

Density functional theory \cite{evans79,evans92} is a primary tool to
treat spatially inhomogeneous systems. For the common reference model
of additive mixtures of hard spheres, the fundamental measures theory
(FMT) \cite{rosenfeld89} is many investigators' current choice. This
approach was extended to cover the AOV model \cite{schmidt00cip},
thereby triggering much further interest in the study of interfacial
properties of this model, see e.g.\ \cite{schmidt04aog} for recent
work on novel effects in sedimentation-diffusion equilibrium. In a
different direction, FMT was recently generalized to hard core {\em
lattice} models \cite{lafuente02,lafuente03diagram,lafuente04prl}
elucidating the very foundation of the FMT approach.

Despite its usefulness for practical applications, the AOV functional
of Ref.\ \cite{schmidt00cip} possesses several deficiencies, that are
absent in the hard sphere FMT (see Ref.\ \cite{schmidt02cip} for a
detailed discussion): i) When applied to one spatial dimension a
spurious phase transition is predicted \cite{schmidt02cip}, that is
absent (as befits a model with short-ranged interactions) in the exact
solution \cite{brader01oned}.  ii) In three dimensions the location of
the bulk fluid-fluid critical point lies at lower polymer fugacity as
compared to the value obtained with simulations. iii) The excess free
energy is a linear functional of the polymer density profile.

While we will not present a full resolution of the above issues,
dealing with a lattice model allows us to go significant steps beyond
the recipe of construction of Ref.\ \cite{schmidt00cip}.  Hence we
consider a simplified version of the AOV model by constraining the
position coordinates to an underlying lattice, for convenience taken
to be of simple cubic symmetry. We are inspired by the fact that
similar lattice models have been proven useful in soft matter research
to make qualitative predictions, like e.g.\ for demixing
\cite{widom67,louis92} and three-phase equilibria \cite{duijneveldt93}.
Also much vital attention has been recently paid to lattice models with
short-ranged attractive interactions to address adsorption in
disordered porous media
\cite{kierlik01,kierlik02,sarkisov02,detcheverry03,woo03}.

We consider the case of small particles sizes, namely such that
particles only repel other particles from nearest neighbor sites. As
the polymers are ideal, each site may be occupied by more than one of
those particles.  We compare two different DFTs for the binary AOV
lattice model, both based on the lattice fundamental measure theory
(LFMT) \cite{lafuente02,lafuente03diagram,lafuente04prl}.  The first
approach allows for multi-occupancy of polymers in the
zero-dimensional limit, and hence constitutes the lattice version of
the (continuum) theory of Ref.\ \cite{schmidt00cip}.  The second
approach relies on the introduction of {\em polymer clusters} as
quasi-species that feature hard-core interactions only. We hence
arrive at an extended model that possesses a ``Highlander''
\cite{onlyOne} property: In an appropriately small cavity, there can
only be one particle. Exploiting further the fact that LFMT can cope
directly with small non-additivities \cite{lafuente02},
we arrive at a theory for the AOV
model that is exact when applied to one dimension; hence it does
correctly predict the absence of phase transitions.  We find that
although the derivations of the two DFTs, as well as their appearance
at first glance differ markedly, the results for bulk phase behavior
in three dimensions are very similar, demonstrating the internal
consistency of FMT.

The paper is organized as follows. In Sec.\ \ref{SECmodel} we define
the lattice AOV model. Sec.\ \ref{SECtheory} is devoted to the
construction of both density functional theories. In Sec.\
\ref{SECresults} results for the bulk phase diagram are presented and
we conclude in Sec.\ \ref{SECconclusions}.

\section{The model}
\label{SECmodel}

\begin{figure}
  \begin{center}
    \includegraphics[width=0.90\columnwidth]{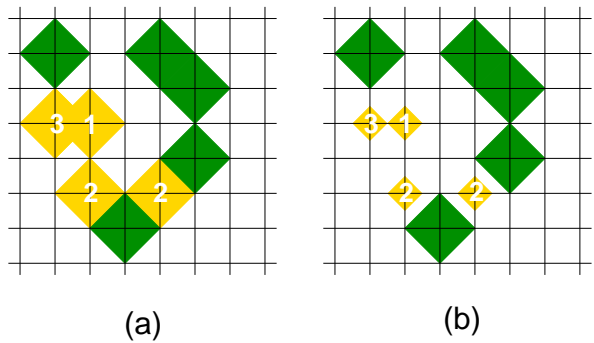}
    \caption{(a) Illustration of the lattice AOV model realized in
      $d=2$ spatial dimensions. Colloidal particles (dark rhombi)
      exclude their site and their nearest neighbor sites to colloids
      and polymers. Polymer particles (light rhombi) exclude colloids
      from their site and their nearest neighbor sites, but can
      overlap freely with other polymers. Integers indicate the number
      of (overlapping) polymers at occupied sites. (b) Equivalent
      non-additive multi-component hard core mixture. Clusters of
      overlapping polymers in (a) are viewed as quasi-particles (light
      rhombi) that exclude their site to other polymer clusters, and
      their site and their nearest neighbor sites to
      colloids. Integers enumerate the different species of polymer
      clusters and correspond to the number of overlapping polymers on
      the lattice sites in (a).  Colloids (dark rhombi) behave as in
      (a).}
\label{FIGmodel}
\end{center}
\end{figure}

We consider a binary mixture of particles representing colloids
(species $c$) and polymers (species $p$) on the $d$-dimensional simple
cubic lattice $\mathbb{Z}^d$. The interaction between colloids is that
of site-exclusion and nearest neighbor exclusion, corresponding to
the pair interaction potential
\begin{equation}
 V_{cc}(\rv) = \begin{cases} \infty & \mbox{if $|\rv| \leq 1$,} \\ 
   0 & {\rm otherwise,} \end{cases}
\end{equation}
where $\rv\in\mathbb{Z}^d$
is the center-center distance between the particles.
Colloids and polymers interact similarly:
\begin{equation}
 V_{cp}(\rv) = \begin{cases} \infty & \mbox{if $|\rv| \leq 1$,} \\
   0 & {\rm otherwise,} \end{cases}
\end{equation}
while polymers are ideal (non-interacting),
\begin{equation}
 V_{pp}(\rv) = 0.
\end{equation}
In essence this is a discretized AOV model with equal-sized
components; see Fig.~\ref{FIGmodel}a for a sketch.

\section{Density functional theories}
\label{SECtheory}

\subsection{Overview of LFMT}
\label{sec:DFToverview}
As is customary in density functional theory \cite{evans79,evans92},
we express the Helmholtz free energy functional as $F=F_{\rm
id}+F_{\rm ex}$, where the free energy of the binary ideal lattice gas
is
\begin{equation}
  F_{\rm id}[\rho_c,\rho_p] = \sum_{i=c,p}\sum_{\rv\in\mathcal{L}}
\rho_i(\rv)[\ln(\rho_i(\rv))-1],
\end{equation}
with $\rho_i(\rv)$ being the occupancy probability of site $\rv$ by
particles of species $i=c,p$, and $\mathcal{L}$ denoting the lattice,
here $\mathcal{L}=\mathbb{Z}^d$. For the excess contribution to the
total free energy, $F_{\rm ex}$, we will in the following use two
different implementations of LFMT. This theory permits to obtain an
approximation to $F_{\rm ex}$ for a lattice model from the exact
solution obtained in finite (and small) sets of lattice sites,
alternatively called nodes
\cite{lafuente02,lafuente03diagram,lafuente04prl}. The essential step
determining the accuracy of the theory is the choice of {\em maximal
cavities} \cite{lafuente04prl}. Those are maximal in the sense that
any further cavity taken into consideration can be obtained as an
intersection of maximal cavities. The common choice (inherited from
the continuum version of the theory \cite{RSLTshort,Tarazona97,
Tarazona00}) for these maximal cavities is to set them equal to
zero-dimensional (0d) cavities, as detailed below. Once chosen, the
cavities uniquely determine the form of the functional under the
requirement that it yields the exact result when evaluated at 0d
density profiles \cite{lafuente04prl}. For a general binary mixture
(with species labeled by $c$ and $p$) the final result possesses the
form
\begin{equation}
F_{\rm ex}[\rho_c,\rho_p]=\sum_{{\cal C}\,\rm cavities}a(\mathcal{C})
F_{\rm ex}^{\mathcal{C}}[\rho_c,\rho_p],
\end{equation}
where the summation runs over the set of maximal cavities and their
nonempty intersections; the $a(\mathcal{C})$ are uniquely determined
integer coefficients \cite{lafuente04prl}, and $F_{\rm ex}^{\mathcal{C}}$
is the {\em exact} excess free energy functional of the system when
confined to cavity $\mathcal{C}$.

\subsection{Multi-occupancy version of DFT}
\label{sec:DFTA}

\begin{figure}
  \begin{center}
    \includegraphics[width=0.65\columnwidth]{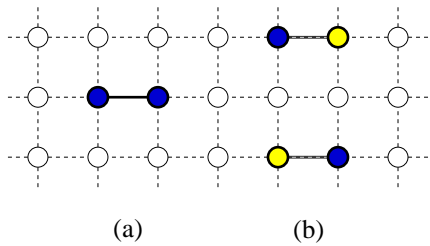}
    \caption{Maximal cavities along the X axis (the ones along the
	Y axis are just 90$^{\circ}$ rotations of these ones) for
	the multi-occupancy (a) and the Highlander (b) DFTs. Dark
	nodes can be occupied either by colloidal or by polymer
	particles; light nodes can only be occupied by colloidal
	particles.}
\label{FIGcavities}
\end{center}
\end{figure}

The definition of 0d cavities for lattice models with only hard-core
interactions is unambiguous: Any set of nodes that can accommodate at
most one particle constitutes a 0d cavity. For the present model,
however, this is problematic: As polymers are ideal, even a single
lattice site can be multiply occupied by polymers, hence no 0d cavity
can contain polymers, and the resulting functional would account for
them only through the ideal part, entirely ignoring the
colloid-polymer interaction. One way out is to relax the definition:
Any set of nodes in which any two particles present {\em necessarily
overlap} constitutes a 0d cavity \cite{schmidt00cip}. Hence, as
opposed to the first definition, the presence of more than one
particle is not excluded a priori. (For mixtures with only hard core
interactions, both definitions are equivalent.)

For the present model it is straightforward to show that according to
the modified (multi-occupancy) definition, any pair of adjacent nodes
on the simple cubic lattice constitutes a maximal 0d cavity 
(Fig.~\ref{FIGcavities}a): Such an
arrangement can either hold one colloid or an arbitrary number of
(overlapping) polymers. Adding an additional node destroys the 0d
character as (at least) two nonoverlapping particles will fit.
Implementing the scheme of Ref.~\onlinecite{lafuente04prl} with this
choice of basic cavities we obtain
\begin{equation}
\beta F_{\rm ex}[\rho_c,\rho_p] =\sum_{\rv}\Phi(\rv),
\end{equation}
with $\beta=1/(k_BT)$, where $T$ is temperature and $k_B$ is the
Boltzmann constant, and the (scaled) free energy density is
\begin{equation}
\begin{split}
  \Phi(\rv) =&\left[\sum_{\alpha=1}^d \phi^{\rm AOV}_0\Big(n^{(\alpha)}_c(\rv),
        n^{(\alpha)}_p(\rv) \Big) \right] \\
    &-(2d-1)\phi^{\rm AOV}_0\Big(\rho_c(\rv),\rho_p(\rv)\Big),
\end{split}
\label{eq:PhiAOV}
\end{equation}
depending on weighted densities
\begin{equation}
  n^{(\alpha)}_i(\rv) \equiv\, \rho_i(\rv)+\rho_i(\rv+\eal),
  \qquad i=c,p,
\end{equation}
where $\eal$ denotes the unit Cartesian vector along the
$\alpha=1,\ldots,d$ axis and
\begin{equation} 
 \phi^{\rm AOV}_0(\ec,\ep)=(1-\ec-\ep)\ln(1-\ec)+\ec
\label{eq:phi0AOV}
\end{equation}
is the excess free energy of a 0d cavity for the AOV model
\cite{schmidt00cip}; we give a derivation of $\phi^{\rm
AOV}_0(\eta_c,\eta_p)$ in the Appendix. Since $\phi^{\rm AOV}_0(\ec,\ep)$
depends linearly on $\ep$ as
\begin{equation}
\phi^{\rm AOV}_0(\ec,\ep)=\phi_0(\ec)+\ep\phi'_0(\ec), 
\end{equation}
where
\begin{equation}
\phi_0(\eta)\equiv\eta+(1-\eta)\ln(1-\eta),
\label{eq:phi0}
\end{equation}
(\ref{eq:PhiAOV}) admits the rewriting
\begin{equation}
\begin{split}
  \Phi =&\left[\sum_{\alpha=1}^d \phi_0(n^{(\alpha)}_c)\right]
    -(2d-1)\phi_0(\rho_c) \\
   &-\rho_p\ln\left(\frac{\prod_{\alpha=1}^d(1-n^{(\alpha)}_c)
   (1-\bar n^{(\alpha)}_c)}{(1-\rho_c)^{2d-1}}\right),
\end{split}
\end{equation}
where, for the sake of notational simplicity, we omit here and in the
following the dependence of $\rho_c,\rho_p, n_c^{(\alpha)}$ on $\rv$
and have introduced the short-hand notation $\bar n_c^{(\alpha)}\equiv
n_c^{(\alpha)}(\rv-\eal)$.

The linear dependence of the functional on $\rho_p(\rv)$ yields a
particularly simple form of the Euler-Lagrange equation for the polymer
density distribution,
\begin{equation}
\rho_p=z_p\frac{\prod_{\alpha=1}^d(1-n^{(\alpha)}_c)
   (1-\bar n^{(\alpha)}_c)}{(1-\rho_c)^{2d-1}},
\label{eq:rhopAOV}
\end{equation}
being explicit in $\rho_p$ (i.e., independent on the right hand side
of $\rho_p$).  Given that we have eliminated $\rho_p(\rv)$ as a
functional of $\rho_c(\rv)$, $\rho_c(\rv\pm\eal)$, $\alpha=1,
\dots,d$, and $z_p(\rv)$, the Legendre-transformed functional
$\beta\Upsilon_{\rm ex}[\rho_c,z_p]=\sum_{\rv}\Phi^{\rm eff}(\rv)$,
with
\begin{equation}
\Phi^{\rm eff}=\rho_p\ln(\rho_p/z_p)-\rho_p+\Phi,
\end{equation}
is an excess free energy functional for the {\em effective}
one-component fluid of colloidal particles interacting through polymer
depletion; $\rho_p(\rv)$ is obtained from (\ref{eq:rhopAOV}). The
strength of the depletion interaction is controlled by $z_p(\rv)$,
which in turn can be determined by an external potential acting on the
polymers and by the polymer fugacity.  (A more detailed account of the
procedure to obtain functionals for effective fluids can be found in
Ref.~\onlinecite{Cuesta99}.)  With the appropriate substitutions, we
obtain
\begin{equation}
\begin{split}
\Phi^{\rm eff} &=\left[\sum_{\alpha=1}^d \phi_0(n^{(\alpha)}_c)\right]
-(2d-1)\phi_0(\rho_c) \\
   &-z_p\frac{\prod_{\alpha=1}^d(1-n^{(\alpha)}_c)
   (1-\bar n^{(\alpha)}_c)}{(1-\rho_c)^{2d-1}}.
\end{split}
\label{eq:effAOV}
\end{equation}
This completes the prescription for the lattice analog of the
continuum AOV functional of Ref.\ \cite{schmidt00cip}.

\subsection{Highlander version of DFT}
\label{sec:DFTB}
\subsubsection{Strategy of polymer clusters as quasi-particles}
Here we elude the presence of several particles inside a cavity and
aim at sticking to the stronger (Highlander) definition: {\em There
can be only one} \cite{onlyOne} (here: particle in the cavity). For
this purpose, polymer ideality constitutes a major problem, as in the
AOV lattice model each site occupied by a polymer can be occupied
simultaneously by an arbitrary number of further polymers. To
circumvent this problem we will hence map the model onto an extended
one: We refer to a piling of $n$ polymers at the same node as a
``polymer cluster'' and treat polymer clusters as quasi-particles of a
set of new species labeled by $n$.  Thus there is a one-to-one
correspondence between each configuration of polymers of the original
model and a corresponding (unique) configuration of polymer {\em
clusters} of the new model. (Colloids are treated as before.)  We
regard polymer clusters as having shapes smaller than the original
polymer, see Fig.\ \ref{FIGmodel}b. As a consequence, polymer clusters
behave as hard core particles with i) site exclusion to other polymer
clusters and ii) site exclusion and nearest neighbor exclusion to
colloids. Clearly, feature ii) is directly inherited from the
colloid-polymer interaction. Feature i) may be unexpected at first
glance. Consider that a site occupied by an $n$-polymer cluster cannot
be {\em simultaneously} occupied by an $m$-polymer cluster. Although
in the original model the site can well be occupied by $m+n$ polymers,
in the extended model such a configuration corresponds to one single
$(m+n)$-polymer cluster. Hence polymer clusters repel like hard cores
do.

Despite being nonadditive, the new model belongs to a class of
nonadditive lattice models which are amenable to LFMT and whose
one-dimensional version was shown to be exact \cite{lafuente02}. The
0d cavities for the new model are nontrivial and consist of two
adjacent nodes available for the colloids, one of which (but only one)
is also available to a quasi-particle (polymer cluster) of any
species (Fig.~\ref{FIGcavities}b).
Clearly, the presence of either one colloidal particle or one
quasi-particle excludes other particles from the cavity, and such
cavities are obviously maximal.

Here we will construct the corresponding free energy functional in two
steps. First, for simplicity, we consider the quasi-particle model
with a single quasi-particle species, $n=1$. Second, we will handle
the full (infinite) number of quasi-particle species.  Subsequently
mapping the result back yields a DFT for the lattice AOV model.

\subsubsection{Binary mixture with one quasi-species}
\label{sec:binaryMixtureWithOneQuasiSpecies}
For the binary mixture of colloids and $1$-polymer clusters, with
density fields $\rho_c(\rv)$ and $\rho_1(\rv)$, respectively,
according to Refs.~\onlinecite{lafuente02,lafuente04prl}, the excess
free energy density $\Phi(\rv)$ is given by
\begin{eqnarray}
\Phi(\rv) &=& \sum_{i=1}^d\left\{
\phi_0\left(n_c^{(\alpha)}(\rv)+\rho_1(\rv)\right) \right.
\nonumber \\
&&\left. +
\phi_0\left(n_c^{(\alpha)}(\rv-\eal)+\rho_1(\rv)\right)-
\phi_0\left(n_c^{(\alpha)}(\rv)\right)
\right\} \nonumber \\
&&- (2d-1)\phi_0\Big(\rho_c(\rv)+\rho_1(\rv)\Big),
\label{eq:binary}
\end{eqnarray}
with $\phi_0(\eta)$ defined in Eq.~(\ref{eq:phi0}). Thus, given a
local fugacity field $z_1(\rv)$, the Euler-Lagrange equation for the
1-polymer cluster density is
\begin{equation}
\rho_1=z_1\frac{\prod_{\alpha=1}^d(1-n_c^{(\alpha)}-\rho_1)
 (1-\bar n_c^{(\alpha)}-\rho_1)}{(1-\rho_c-\rho_1)^{2d-1}}.
\label{eq:ELrhop}
\end{equation}
where again the dependence on $\rv$ has been omitted.  Eq.\
(\ref{eq:ELrhop}) is an implicit algebraic equation for $\rho_1(\rv)$,
so it determines $\rho_1(\rv)$ as a functional of $\rho_c(\rv)$ and
$z_1(\rv)$. This permits to obtain, as in the previous case, a
functional $\beta\Upsilon_{\rm ex}[\rho_c,z_1]$ for the effective
one-component fluid.

\subsubsection{Mixture with an infinite number of quasi-species}

Incorporating the infinite number of quasi-species amounts to first
correctly modifying the ideal free energy functional to $\beta F_{\rm
id}[\rho_c,\{\rho_n\}]=\sum_{\rv}\Phi_{\rm id}(\rv)$, with
\begin{equation}
\begin{split}
\Phi_{\rm id}(\rv)=&\,
\rho_c(\rv)\ln(\rho_c(\rv))-\rho_c(\rv) \\
&+ \sum_{n=1}^{\infty}\Big[\rho_n(\rv)\ln\Big({\cal V}_n\rho_n(\rv)
\Big) -\rho_n(\rv)\Big],
\end{split}
\label{eq:phiideal}
\end{equation}
where $\rho_n(\rv)$ denotes the density profile of the quasi-species
of $n$-polymer clusters, and ${\cal V}_n$ a thermal ``volume''
accounting for the internal partition function of the polymers inside
the cluster. (It is not hard to guess that $\mathcal{V}_n=n!$;
nevertheless, this result will emerge explicitly from the subsequent
analysis.) Secondly, we use again LFMT for the excess part of the free
energy. The necessary modification over the binary case (Sec.\
\ref{sec:binaryMixtureWithOneQuasiSpecies}) in order to include the
infinite number of quasi-species is actually very simple
\cite{lafuente02}: we replace $\rho_1(\rv)$ in (\ref{eq:binary}) by
the total density of polymer clusters (pc),
\begin{equation}
\rhocluster(\rv)\equiv\sum_{n=1}^{\infty}\rho_n(\rv).
\label{eq:rhoinf}
\end{equation}

As our aim is to describe the lattice AOV model, we have to ``project''
the quasi-species back to polymers. Clearly, the total cluster density
and the polymer density are related via
\begin{equation}
\rho_p(\rv)=\sum_{n=1}^{\infty}n\rho_n(\rv).
\label{eq:constraint}
\end{equation}
For convenience we will again resort to an effective one-component
description. Hence we transform to the semi-grand potential
\begin{equation}
\beta\Upsilon[\rho_c,z_p]=\min_{\{\rho_n\}}\sum_{\rv}\Big(\Phi_{\rm id}(\rv)+
\Phi(\rv)-\rho_p(\rv)\ln z_p(\rv)\Big),
\label{eq:upsilon}
\end{equation}
where $\Phi_{\rm id}(\rv)$ is given by (\ref{eq:phiideal}), and 
$\Phi(\rv)$ by (\ref{eq:binary}) with $\rho_1(\rv)$ replaced by 
$\rhocluster(\rv)$. With the same notation as in (\ref{eq:ELrhop}),
performing the minimization in (\ref{eq:upsilon}) yields
\begin{equation}
\rho_n=\frac{z_p^n}{\mathcal{V}_n}\frac{\prod_{\alpha=1}^d
  (1-n_c^{(\alpha)}-\rhocluster)(1-\bar n_c^{(\alpha)}-\rhocluster)}
  {(1-\rho_c-\rhocluster)^{2d-1}},
\label{eq:ELrhon}
\end{equation}
which using (\ref{eq:rhoinf}) leads to
\begin{equation}
\rhocluster=\zeta(z_p)\frac{\prod_{\alpha=1}^d(1-n_c^{(\alpha)}-\rhocluster)
(1-\bar n_c^{(\alpha)}-\rhocluster)}{(1-\rho_c-\rhocluster)^{2d-1}},
\label{eq:ELrhoinf}
\end{equation}
with $\zeta(z_p)\equiv\sum_{n=1}^{\infty}(z_p^n/\mathcal{V}_n)$.
Also, from Eqs.~(\ref{eq:constraint}), (\ref{eq:ELrhon}) and
(\ref{eq:ELrhoinf}) we get
\begin{equation}
\rho_p=z_p\frac{\zeta'(z_p)}{\zeta(z_p)}\rhocluster,
\label{eq:rhoprhoinf}
\end{equation}
establishing a simple proportionality between $\rho_p(\rv)$ and
$\rhocluster(\rv)$.

The function $\zeta(z_p)$ can be easily obtained realizing that it is
{\em independent} of all densities appearing in (\ref{eq:ELrhoinf}),
rather it depends solely on $z_p(\rv)$. Thus if we particularize
(\ref{eq:ELrhoinf}), e.g.\ for $\rho_c(\rv)=0$, we obtain the simple
relationship
\begin{equation}
\rhocluster=\zeta(z_p)(1-\rhocluster),
\end{equation}
from which
\begin{equation}
\rhocluster=\frac{\zeta(z_p)}{1+\zeta(z_p)}.
\end{equation}
Then, according to Eq.~(\ref{eq:rhoprhoinf}),
\begin{equation}
\rho_p=z_p\frac{\zeta'(z_p)}{1+\zeta(z_p)}
=z_p\Big(\ln(1+\zeta)\Big)',
\end{equation}
and since in the absence of colloidal particles, the polymers form
an ideal gas (hence $\rho_p=z_p$), we get
\begin{equation}
\Big(\ln(1+\zeta)\Big)'=1.
\end{equation}
The solution to this equation satisfying $\zeta(0)=0$ is
\begin{equation}
\zeta(z_p)=e^{z_p}-1=\sum_{n=1}^{\infty}\frac{z^n}{n!},
\label{eq:zinfzp}
\end{equation}
providing the confirmation of the relationship $\mathcal{V}_n=n!$.

\subsubsection{Functional for the lattice AOV model}

Returning to Eq.~(\ref{eq:ELrhoinf}), we observe that this equation
for $\rhocluster(\rv)$ is formally equivalent to Eq.~(\ref{eq:ELrhop})
for $\rho_1(\rv)$ with $z_1(\rv)$ replaced by the
$\zeta\big(z_p(\rv)\big)$ given by (\ref{eq:zinfzp}).  Recalling the
origin of Eq.~(\ref{eq:ELrhop}), this means that we can describe the
original AOV model in terms of the free energy functional $\beta
F[\rho_c,\rhocluster]=\sum_{\rv}\left\{ \Phi_{\rm
id}(\rv)+\Phi(\rv)\right\}$, with
\begin{eqnarray}
\Phi_{\rm id} &=& \rho_c(\ln\rho_c-1)+
\rhocluster(\ln \rhocluster-1) \nonumber \\
\Phi &=& \sum_{i=1}^d\left [
\phi_0(n_c^{(\alpha)}+\rhocluster) +
\phi_0(\bar n_c^{(\alpha)}+\rhocluster) \right. \nonumber \\
&&-\left.\phi_0(n_c^{(\alpha)})\right ]
- (2d-1)\phi_0(\rho_c+\rhocluster),
\end{eqnarray}
with the fugacity for the total density of polymer clusters,
$\rhocluster(\rv)$, given by
\begin{equation}
\zeta(\rv)=e^{z_p(\rv)}-1,
\label{eq:zinfzp2}
\end{equation}
and the polymer density obtained from $\rhocluster(\rv)$ as
\begin{equation}
\rho_p(\rv)=\frac{z_p(\rv)}{1-e^{-z_p(\rv)}}\,\rhocluster(\rv).
\label{eq:rhoprhoinf2}
\end{equation}
This completes the prescription of the Highlander functional for the
AOV model.

As a final note, we can alternatively obtain the effective functional
$\beta\Upsilon_{\rm ex}[\rho_c,z_p]=\sum_{\rv}\Phi^{\rm eff}(\rv)$,
which, after a few algebraic manipulations, can be written as
\begin{equation}
\begin{split}
\Phi^{\rm eff}=&\,
\rho_c\ln\rho_c-(2d-1)(1-\rho_c)
\ln(1-\rho_c-\rhocluster) \\
&+\sum_{\alpha=1}^d\big[(1-n_c^{\alpha})\ln(1-n_c^{\alpha}-\rhocluster) \\
&+(1-\bar n_c^{\alpha})\ln(1-\bar n_c^{\alpha}-\rhocluster) \\
&-(1-n_c^{\alpha})\ln(1-n_c^{\alpha}) \big],
\end{split}
\label{eq:finalPhieff}
\end{equation}
with $\rhocluster(\rv)$ being the solution of (\ref{eq:ELrhoinf}).

\subsection{Relationship between both DFTs}

Assuming small polymer fugacity, $z_p(\rv)\ll 1$, for the Highlander
DFT, from Eqs.~(\ref{eq:zinfzp2}) and (\ref{eq:rhoprhoinf2}) it
follows that $\zeta(\rv)\approx z_p(\rv)$ and $\rhocluster(\rv)\approx
\rho_p(\rv)$. The functional is then approximately given by
Eq.~(\ref{eq:binary}). On the other hand, small fugacities imply small
densities, so we can expand (\ref{eq:binary}) to linear order in
$\rho_p(\rv)$. Doing so and taking into account that the sum over
$\rv$ allows us to shift the arguments of the functions, we realize
that the expansion is precisely functional (\ref{eq:PhiAOV}). So both
DFTs coincide at low polymer fugacities.

\section{Results}
\label{SECresults}

As an application we consider the bulk phase diagram for the lattice
AOV model and compare the predictions from both DFTs. Guided by its
continuum counterpart, we expect the phase behavior of the lattice
model to be determined by the interplay of hard core colloid-colloid
repulsion and the (short-ranged) colloid-colloid attraction induced by
polymer depletion. Thus a clear framework to study the model is the
effective one-component description, cf.~Eqs.~(\ref{eq:effAOV}) and
(\ref{eq:finalPhieff}). As in other (one-component) models with hard
core repulsion and short-range attraction, both condensation
(corresponding to demixing from the viewpoint of the binary mixture)
and freezing may occur. We can study the former via a convexity
analysis of the effective (semi-grand) free energy of a uniform fluid,
and the latter by considering spatially inhomogeneous density
distributions characteristic for crystalline states. Here the
candidate is a face-centered cubic (fcc) structure that constitutes
the close packed state for the colloids. In the following we will
carry out this program for either DFT.

\subsection{Thermodynamics from the multi-occupancy DFT}
\label{subsec:thermomulti}
As an appropriate thermodynamic potential we consider a (scaled)
semi-grand free energy per volume, $Y\equiv 
\rho_c[\ln(\rho_c)-1]+\beta\Upsilon_{\rm ex}/V$. For a uniform fluid at
given polymer fugacity $z_p$, according to (\ref{eq:effAOV}) this is
given by
\begin{equation}
\begin{split}
Y=&\,\frac{\eta_c}{2}\ln\left(\frac{\eta_c}{2}\right)+d(1-\eta_c)\ln(1-\eta_c) \\
&-(2d-1)\left(1-\frac{\eta_c}{2}\right)
\ln\left(1-\frac{\eta_c}{2}\right) \\
&-z_p\frac{(1-\eta_c)^{2d}}{\left[1-(\eta_c/2)\right]^{2d-1}},
\end{split}
\label{eq:YunifA}
\end{equation}
where $\eta_c=2\rho_c$ is the colloid packing fraction. Convexity
implies $Y''(\eta_c)>0$, so the spinodal condition, $Y''(\eta_c)=0$,
yields
\begin{equation}
\eta^{r, \spin}_p(\eta_c)=\frac{2\left(1-\frac{\eta_c}{2}\right)^{2d}
[1+(d-1)\eta_c]}{d(2d-1)\eta_c(1-\eta_c)^{2d-1}},
\label{eq:condensA}
\end{equation}
where we have used the polymer reservoir packing fraction, $\eta^{\rm
r}_p=2z_p$, as an alternative thermodynamic variable. The minimum of
this curve determines the critical point, given by one of the roots of
the cubic polynomial
\begin{equation}
(d-1)(\eta_c^\crit)^3+2(d-1)^2(\eta_c^\crit)^2+(2d+1)\eta_c^\crit-2=0,
\end{equation}
and the corresponding value $\eta_p^{r,\crit}$ obtained from
(\ref{eq:condensA}). For $d=3$ the critical point is at 
$\eta_c^\crit=\sqrt{3/2}-1\approx 0.225$
(see Fig~\ref{FIGbulkPhaseDiagram}). The liquid-gas
binodal can be obtained from (\ref{eq:YunifA}) via a double tangent
construction, in practice carried out numerically.

For crystals we are guided by the close-packed state of the colloids,
and distinguish two fcc sublattices of the underlying simple cubic
lattice, each formed by the nearest neighbor nodes of the other.  The
colloid density at nodes of sublattices $a$ and $b$ are denoted by
$\rho_a$ and $\rho_b$, respectively. Hence the average colloid packing
fraction is
\begin{equation}
 \eta_c=\rho_a+\rho_b.
 \label{eq:crystalDensityConstraint}
\end{equation} 
The semi-grand potential is given by
\begin{equation}
\begin{split}
Y=&\, \frac{\rho_a}{2}\ln\rho_a+\frac{\rho_b}{2}\ln\rho_b
+d(1-\eta_c)\ln(1-\eta_c) \\
&-\left(d-{\textstyle\frac{1}{2}}\right)(1-\rho_a)\ln(1-\rho_a) \\
&-\left(d-{\textstyle\frac{1}{2}}\right)(1-\rho_b)\ln(1-\rho_b) \\
&-\frac{z_p}{2}\left[\frac{(1-\eta_c)^{2d}}{(1-\rho_a)^{2d-1}}
+\frac{(1-\eta_c)^{2d}}{(1-\rho_b)^{2d-1}}\right].
\end{split}
\end{equation}
The equilibrium density distribution for given values of $\eta_c$ can
be obtained by minimizing $Y$ with respect to $\rho_a$ and $\rho_b$
under the constraint (\ref{eq:crystalDensityConstraint}), leading to
the condition
\begin{equation}
\begin{split}
\rho_a & (1-\rho_a)^{2d-1}\exp\left(-(2d-1)z_p\left(\frac{1-\eta_c}{1-
\rho_a}\right)^{2d}\right) \\
&=\rho_b(1-\rho_b)^{2d-1}\exp\left(-(2d-1)z_p\left(\frac{1-\eta_c}{1-
\rho_b}\right)^{2d}\right).
\end{split}
\end{equation}
This equation possesses one solution characteristic for fluid states,
$\rho_a=\rho_b=\eta_c/2$, which is locally stable as long as it
minimizes $Y(\rho_a,\rho_b)$, i.e.\ up to the value of $\eta_c$ at
which
\begin{equation}
\frac{\partial^2}{\partial\rho_a^2}Y(\rho_a,\eta_c-\rho_a)\bigg|_{\rho_a=
\eta_c/2}=0,
\label{eq:freezingcondition}
\end{equation}
defining the freezing spinodal, that we can explicitly obtain as
\begin{equation}
\eta^{\rm r,fr}_p(\eta_c)=\frac{2\left(1-\frac{\eta_c}{2}\right)^d
(1-d\eta_c)}{d(2d-1)\eta_c(1-\eta_c)^{2d}}.
\label{eq:freezingA}
\end{equation}

If freezing is second order, Eq.\ (\ref{eq:freezingA}) directly gives
the transition line in the phase diagram. In the case of a first-order
transition again a numerical double tangent construction is required
to obtain the coexistence densities.

\subsection{Thermodynamics from the Highlander DFT}
In the fluid phase we obtain the semi-grand potential
\begin{equation}
\begin{split}
Y=&\,\frac{\eta_c}{2}\ln\left(\frac{\eta_c}{2}\right)-
(2d-1)\left(1-\frac{\eta_c}{2}\right)
\ln\left(1-\frac{\eta_c}{2}-\rhocluster\right) \\
&+d(1-\eta_c)\ln\left(\frac{(1-\eta_c-\rhocluster)^2}{1-\eta_c}\right),
\end{split}
\end{equation}
with $\rhocluster$ being the solution of
\begin{equation}
\left(1-\frac{\eta_c}{2}-\rhocluster\right)^{2d-1}\rhocluster=
\zeta(z_p)(1-\eta_c-\rhocluster)^{2d}.
\label{eq:t-eta0}
\end{equation}
Again, the fluid-fluid spinodal can be determined by $Y''(\eta_c)=0$;
necessary expressions for $\rhocluster'(\eta_c)$ and
$\rhocluster''(\eta_c)$ can be obtained implicitly from 
(\ref{eq:t-eta0}). The spinodal polymer cluster density is
\begin{equation}
\rhocluster^{\spin}(\eta_c)=
\frac{(1-\eta_c)\big[1+(d-1)\eta_c\big]}{1+(d-1)(2d+1)\eta_c},
\label{eq:spin1}
\end{equation}
which inserted into (\ref{eq:t-eta0}) gives the liquid-gas spinodal
\begin{equation}
\begin{split}
\eta_p^{\rm r,spin}(\eta_c)=&\,2\ln\Bigg(1+
\frac{2(2d-1)^{2d-1}}{\big[4d(d-1)\big]^{2d}}
\big[1+(d-1)\eta_c\big] \\
&\times 
\frac{\big[2d-1-(d-1)\eta_c\big]^{2d-1}}{\eta_c(1-\eta_c)^{2d-1}}
\Bigg),
\end{split}
\end{equation}
[notice that $\eta_p^{\rm r}=2\ln(1+\zeta)$].
The critical point is at the minimum value of $\eta_p^{\rm r, spin}$
versus $\eta_c$, which we can obtain analytically as
\begin{equation}
  \eta_c^\crit=\frac{1-d(d+1)+d\sqrt{2-6d+5d^2}}{(d-1)^2(2d+1)}.
  \label{eq:criticalPointHighlander}
\end{equation}

In order to treat crystalline states we use the same density
parameterization as in the treatment with the multi-occupancy version,
with $\rho_a$ and $\rho_b$ being the densities of two sublattices, and
$\eta_c=\rho_a+\rho_b$. The semi-grand potential is obtained as
\begin{equation}
\begin{split}
Y =&\,\frac{\rho_a}{2}\ln\rho_a+\frac{\rho_b}{2}\ln\rho_b \\
&-(d-{\textstyle \frac{1}{2}})(1-\rho_a)\ln(1-\rho_a-\rhocluster^{(a)}) \\
&-(d-{\textstyle \frac{1}{2}})(1-\rho_b)\ln(1-\rho_b-\rhocluster^{(b)}) \\
&+d(1-\eta_c)\ln\left(\frac{(1-\eta_c-\rhocluster^{(a)})
  (1-\eta_c-\rhocluster^{(b)})}{1-\eta_c}\right),
\end{split}
\end{equation}
and $\rhocluster^{(\alpha)}$, $\alpha=a,b$, fulfill
\begin{equation}
\rhocluster^{(\alpha)}(1-\rho_{\alpha}-\rhocluster^{(\alpha)})^{2d-1}=
 \zeta(z_p)(1-\eta_c-\rhocluster^{(\alpha)})^{2d}.
\label{eq:t-eta}
\end{equation}
Minimizing $Y(\rho_a,\eta_c-\rho_a)$ with respect to $\rho_a$ leads to
\begin{equation}
\rho_a\big(1-\rho_a-\rhocluster^{(a)})^{2d-1}=\rho_b
\big(1-\rho_b-\rhocluster^{(b)})^{2d-1}.
\label{eq:minimum}
\end{equation}
We obtain $\partial \rhocluster^{(\alpha)}/\partial\rho_{\alpha}$
through implicit differentiation in (\ref{eq:t-eta}).

As in section \ref{subsec:thermomulti}, the freezing spinodal is
obtained through Eq.~(\ref{eq:freezingcondition}), yielding
\begin{equation}
\rhocluster^{(a)}\Big|_{\rho_a=\rho_b=\eta_c/2}
=\frac{(1-\eta_c)(1-d\eta_c)}{1+d(2d-3)\eta_c},
\end{equation}
which combined with (\ref{eq:t-eta}) gives the analytic result
\begin{equation}
\begin{split}
\eta_p^{\rm r,fr}(\eta_c) &=2\ln\Bigg[1+
\frac{2(2d-1)^{2d-1}}{\big[4d(d-1)\big]^{2d}} \\
&\times \frac{(2d-1-d\eta_c)^{2d-1}}{\eta_c(1-\eta_c)^{2d-1}}
\big(1-d\eta_c\big)\Bigg].
\end{split}
\label{eq:freezingspinodal}
\end{equation}

\subsection{Phase behavior in one dimension}
It is clear from a general argument applicable to one-dimensional
systems with short-ranged forces that the lattice AOV model does not
exhibit a phase transition for $d=1$. In this case the Highlander
functional (\ref{eq:binary}), with $\rho_p(\rv)$ replaced by
$\rhocluster(\rv)$, is exact \cite{lafuente02}. (Note that the mapping
of the polymers to hard-core polymer clusters, to which we apply the
theory of \cite{lafuente02}, is an exact transformation.)  Hence an
immediate consequence is that the Highlander version correctly
predicts the absence of phase transitions.  We can recover this result
explicitly: the position of the liquid-gas critical point, obtained
via (\ref{eq:criticalPointHighlander}), is $\eta_c^\crit=1/2$,
$\eta_p^{\rm r}=\infty$, consistent with the absence of the
transition. The freezing spinodal, (\ref{eq:freezingspinodal}),
reduces to $\eta_p^{\rm r} =\infty$, again reflecting the absence of a
phase transitions.  The multi-occupancy version, however, incorrectly
predicts both condensation and freezing, see Eqs.~(\ref{eq:condensA})
and (\ref{eq:freezingA}). The one fluid phase is below (in $\eta_p^r$)
the liquid-gas critical point located at $\eta_c^\crit=2/3$,
$\eta_p^{r,\crit}=4$.

\subsection{Phase behavior in three dimensions}

\begin{figure}
  \begin{center}
    \includegraphics[width=\columnwidth]{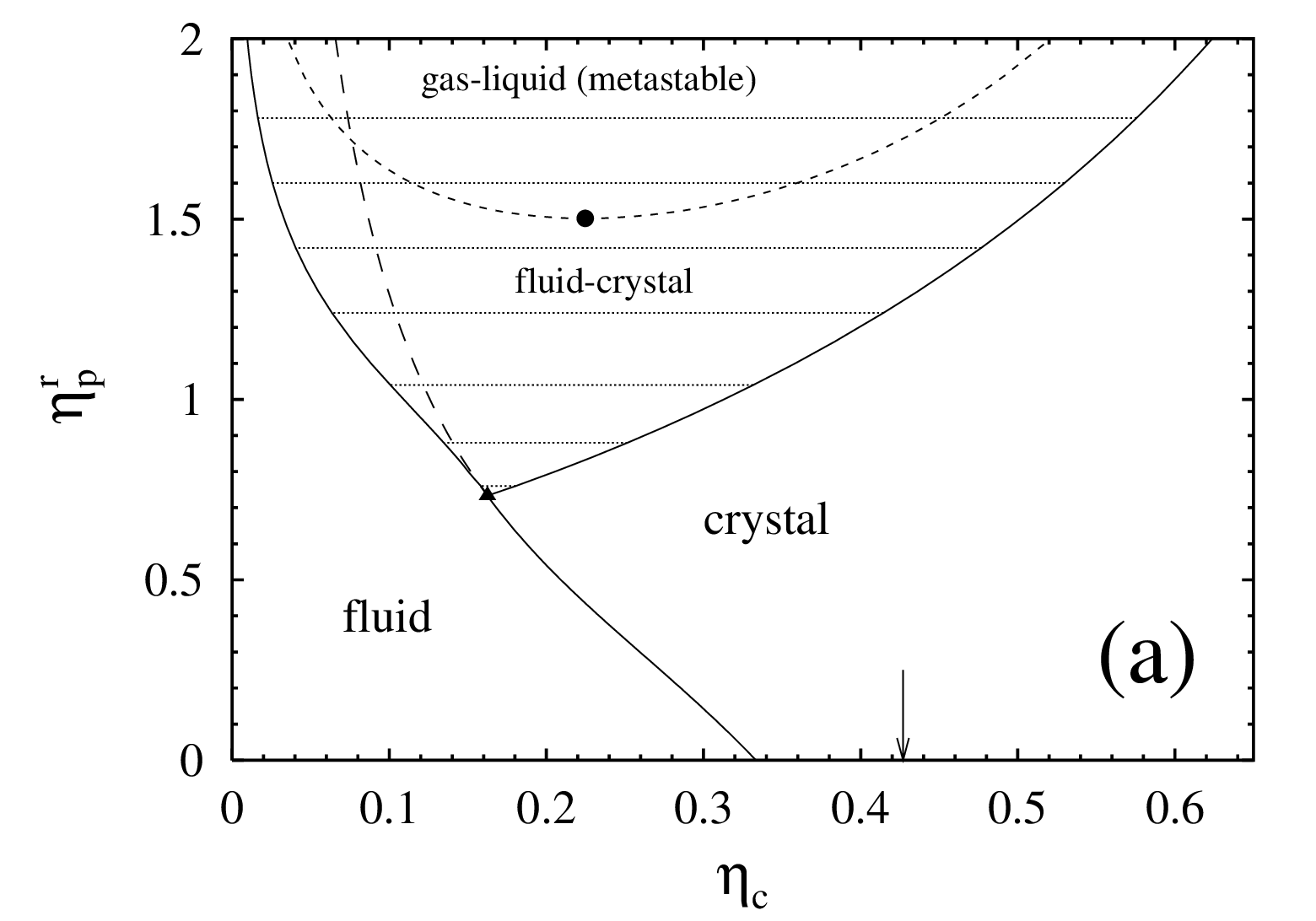}
    \includegraphics[width=\columnwidth]{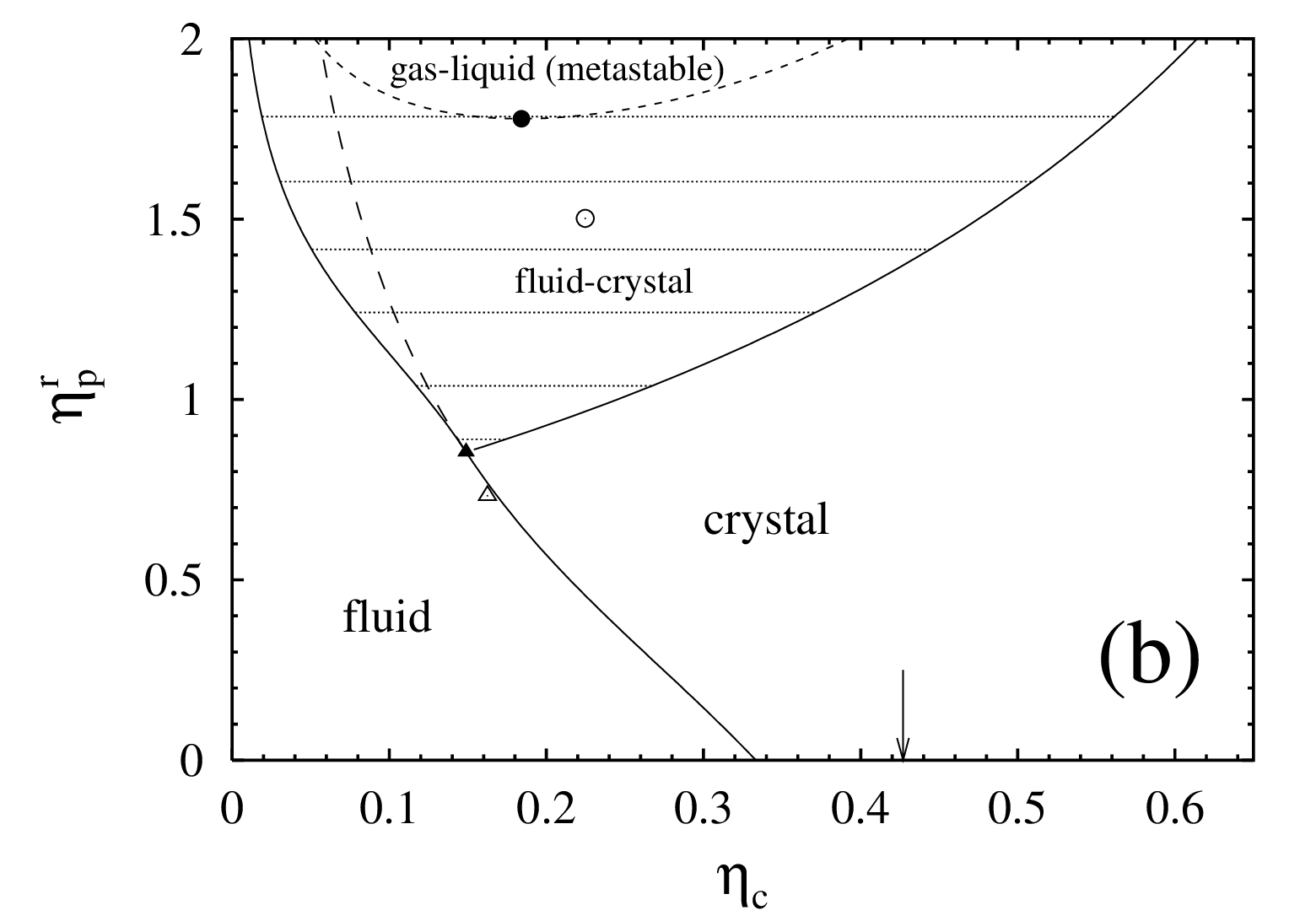}
    \caption{Bulk phase diagram of the lattice AOV model as a function
      of colloid packing fraction, $\eta_c$, and polymer reservoir
      packing fraction, $\eta_p^{\rm r}$. (a) Result from the
      multi-occupancy DFT. For small $\eta_p^{\rm r}$, below the
      tricritical point (triangle), there is a continuous freezing
      transition into an fcc crystal; its location for $\eta_p^{\rm
      r}=0$ is at a smaller value of $\eta_c$ as compared to the
      result by Gaunt \cite{gaunt67} (arrow). Above the tricritical
      point freezing is discontinuous; coexistence is along horizontal
      tielines (dotted lines) between the fluid and crystal branches
      of the binodal (solid lines). Also shown is the freezing
      spinodal (long-dashed line) where the fluid phase loses its
      metastability upon increasing $\eta_c$.  The gas-liquid binodal
      (short-dashed line) ending in a lower critical point (filled
      dot) is metastable with respect to freezing. (b) The same as
      (a), but obtained from the Highlander DFT. For comparison the
      tricritical point (open triangle) and the metastable gas-liquid
      critical point (open dot) obtained from the multi-occupancy DFT
      are shown.}
\label{FIGbulkPhaseDiagram}
\end{center}
\end{figure}

We display the result for the full phase diagram from either DFT in
Fig.~\ref{FIGbulkPhaseDiagram}, plotted as a function of colloid
packing fraction, $\eta_c$, and polymer reservoir packing fraction,
$\eta_p^r$. For $\eta_p^r=0$ the system is a pure (colloid) hard core
lattice gas with nearest-neighbor exclusion. Both DFTs reduce to the
(same) LFMT that predicts a continuous freezing transition at a
(colloid) packing fraction of $\eta_c=1/3$. This is considerably lower
than the value obtained from Pad\'e approximants \cite{gaunt67},
$\eta_c^{\crit}\approx 0.43$.

Increasing $\eta_p^r$ leads to a shift of the transition to smaller
values of $\eta_c$, which can be attributed to polymers substituting
colloids on crystal lattice positions and hence decreasing the colloid
density at the transition. At a threshold value of $\eta_p^r$ the
transition becomes first order and a density gap opens up. The
location of this tricritical point differs somewhat in both
treatments, being located at a larger values of $\eta_p^r$ in the
Highlander DFT. Upon increasing $\eta_p^r$ further, the coexistence density
gap increases strongly. The liquid-gas transition is found to be
metastable with respect to freezing.  (The liquid-gas binodal obtained
from the Highlander version is again located at higher values of
$\eta_p^r$ as compared to the multi-occupancy version.)

The occurrence of a tricritical point for the freezing transition is a
peculiarity of the lattice model absent in the continuum version,
where freezing is first order for all $\eta_p^r$
\cite{dijkstra99}. The phase behavior of the continuum AOV model
depends sensitively on the polymer-to-colloid size ratio,
$q=\sigma_p/\sigma_c$, where $\sigma_p$ and $\sigma_c$ is the polymer
and colloid diameter, respectively. The phase diagram obtained for
small values of $q\sim 0.1$ indeed roughly resembles the topology of
the phase diagram that we find for the lattice AOV model, provided
$\eta_p^r$ is sufficiently high. For small values of $\eta_p^r$ the
broad coexistence region smoothly approaches the hard sphere
crystal-fluid coexistence densities in the continuum model (without an
intervening tricritical point).

Given the fact that we do find the correct (hence also experimentally
observed) fcc crystalline structure, we hence conclude that the
current model is very suitable for the study of inhomogeneous
situations at high $\eta_p^r$, where a dilute colloidal gas and a
dense crystal are the relevant bulk states.

\section{Conclusions}
\label{SECconclusions}

We have carried out a detailed comparison between two density
functional theories for a lattice AOV model consisting of a binary
mixture of colloidal particles and polymers both with position
coordinates on a simple cubic lattice. The (pair) interactions are
such that colloids exclude their site and nearest neighbors to both
colloids and polymers, and polymers do the same with colloids, but do
not interact with other polymers. Relying on the LFMT concept we have
obtained two density functionals for arbitrary space dimension $d$ by
starting with the zero-dimensional Statistical Mechanics of the model.
The multi-occupancy version is an analog of the DFT for the continuum
AOV model \cite{schmidt00cip}. The Highlander version exploits the
possibility of mapping the lattice model to a cluster model that
features hard core interaction only. Both DFTs are exact for $d=0$ by
construction. The Highlander version is also exact in $d=1$, where the
multi-occupancy version incorrectly gives a phase transition.

The predictions for the phase diagram for $d=3$ from both approaches
are very similar, and we have argued that the topology resembles that
of real colloid-polymer mixtures for small polymer-to-colloid size
ratios, where the liquid-gas transition is metastable with respect to
freezing into an fcc colloid crystal and the coexistence density gap
is very wide for large values of polymer fugacity.  We hence conclude
that both the model and DFT approaches are well suited to study
properties of inhomongeneous situations in colloid-polymer mixtures
driven by dilute colloidal fluid and dense colloidal crystal phases.

\acknowledgments

This work is supported by projects no.~BFM2003-0180 of the Direcci\'on
General de Investigaci\'on (Ministerio de Ciencia y Tecnolog\'{\i}a,
Spain) and the SFB-TR6 ``Colloidal dispersions in external fields'' of
the German Science Foundation (Deutsche Forschungsgemeinschaft).

\bigskip

\section*{APPENDIX}

For notational simplicity, we use ${\bf 1}\equiv\rv$ and ${\bf
2}\equiv\rv+\eal$. If $z_i({\bf x})$ denotes the local fugacity at
node ${\bf x}(={\bf 1},{\bf 2})$ of species $i(=c,p)$, then the grand
partition function of a maximal 0d cavity will be
\begin{equation}
\Xi_0=e^{z_p({\bf 1})+z_p({\bf 2})}+z_c({\bf 1})+z_c({\bf 2}).
\label{eq:Xi0}
\end{equation}
Since $\rho_i({\bf x})=z_i({\bf x})\delta\ln\Xi_0/\delta z_i({\bf x})$,
from the equation above
\begin{eqnarray}
\rho_c({\bf x}) &=& \frac{z_c({\bf x})}{\Xi_0},
\label{eq:rhocx} \\
\rho_p({\bf x}) &=& \frac{z_p({\bf x})e^{z_p({\bf 1})+z_p({\bf 2})}}{\Xi_0}.
\label{eq:rhopx}
\end{eqnarray}
From (\ref{eq:rhocx}),
\begin{equation}
\eta_c\equiv\rho_c({\bf 1})+\rho_c({\bf 2})=\frac{z_c({\bf 1})+
z_c({\bf 2})}{\Xi_0},
\end{equation}
so eliminating $z_c({\bf 1})+z_c({\bf 2})$ in (\ref{eq:Xi0}) we
obtain
\begin{equation}
e^{z_p({\bf 1})+z_p({\bf 2})}=\Xi_0(1-\eta_c).
\label{eq:expzp}
\end{equation}
With this Eq.~(\ref{eq:rhopx}) becomes
\begin{equation}
\rho_p({\bf x})=z_p({\bf x})(1-\eta_c),
\label{eq:newrhopx}
\end{equation}
what allows us to obtain from (\ref{eq:expzp})
\begin{equation}
\ln\Xi_0=\frac{\eta_p}{1-\eta_c}-\ln(1-\eta_c)
\label{eq:lnXi0}
\end{equation}
(with the obvious notation $\eta_p\equiv\rho_p({\bf 1})+\rho_p({\bf 2})$).

Now, as $\phi^{\rm AOV}_0=\beta F_{\rm ex}[\rho_c,\rho_p]$ in this cavity,
\begin{equation}
\phi^{\rm AOV}_0=-\ln\Xi_0+\sum_{i,{\bf x}}
\rho_i({\bf x})\ln\Big(z_i({\bf x})/\rho_i({\bf x})\Big),
\end{equation}
and using (\ref{eq:rhocx}), (\ref{eq:newrhopx}) and (\ref{eq:lnXi0})
we finally obtain (\ref{eq:phi0AOV}).

\vfill

\bibliographystyle{apsrev} 
\bibliography{FMF}

\end{document}